\documentclass[12pt]{article}
\usepackage{latexsym,amsmath,amssymb,amsbsy,graphicx}
\usepackage[utf8]{inputenc}
\usepackage[T2A]{fontenc}
\usepackage[space]{cite} 

\makeatletter\def\@biblabel#1{\hfill#1.}\makeatother

\allowdisplaybreaks
\begin{document}


\noindent\begin{minipage}{\textwidth}
\begin{center}

{\Large{ASTRONOMY, ASTROPHYSICS, AND COSMOLOGY}}\\[20pt]

{\Large{Sky background accounting in spectral infrared observations 
of extended objects at the Caucasus Mountain Observatory of the
SAI MSU}}\\[9pt]

{\large Gusev A.\,S.$^{1a}$, Tatarnikov A.\,M.$^{1,2}$, Zheltoukhov S.\,G.$^{1}$, Kirsanova M.\,S.$^{3}$}\\[6pt]

\textit {$^1$Sternberg Astronomical Institute, Lomonosov Moscow State University, Moscow 119234, Russia.}\\
\textit {$^2$Faculty of Physics, Lomonosov Moscow State University, Moscow 119991, Russia.}\\
\textit {$^3$Institute of astronomy of Russian Academy of Sciences, Moscow 119017, Russia.}\\
\textit {E-mail: $^a$gusev@sai.msu.ru}\\

Received 16.02.2026, revised 06.04.2026, accepted 08.04.2026.

\end{center}

\begin{abstract}
The Caucasus Mountain Observatory of the Sternberg Astronomical Institute of Moscow State University is the only one in Russia and one of the few in the world where is possible to obtain spectral data in the near-infrared (IR) range at $\lambda=1-2.5~\mu$m. However, there is a problem of processing the spectra of extended objects, the angular dimensions of which exceed the length of the slit (4.5~arcmin). Obtaining additional spectra of the sky in the immediate vicinity of such objects does not solve the problem, since bright atmospheric hydroxyl lines at $\lambda\sim2~\mu$m change their intensity significantly over a time shorter than the exposure time of a single frame. We have developed a technique that allows us to correctly account for and exclude the contribution of variable atmospheric lines in the spectra of extended objects. This technique has been successfully tested in spectroscopic studies of the star-forming region NGC~7538 (S158) in our Galaxy.
\end{abstract}

\vspace{1pt}\par
\end{minipage}



\section{Introduction}\label{intro}

The Astronomical Educational and Scientific Complex ''Caucasian Mountain Observatory of the SAI MSU'' (CMO~\cite{shatsky2020}) is the most modern astronomical observatory in Russia, allowing photometric, spectral, and speckle-polarimetric observations in the optical and infrared ranges. The observatory's flagship, the 2.5-meter Ritchey-Chr\'{e}tien telescope, is equipped with an infrared camera-spectrograph, AstroNIRCam~\cite{nadjip2017}, unique in Russia and one of the few in the world, allowing for photometric, spectrophotometric~\cite{tatarnikov2023}, and spectroscopic observations~\cite{zheltoukhov2020} in the range from 1.04 to 2.45~$\mu$m.

Despite the fact that the IR data obtained with the CMO are somewhat inferior in quality to the results obtained with specialized IR telescopes and, above all, the James Webb Space Telescope (JWST)~\cite{jakobsen2022,boker2022}, the SAI team manages to obtain unique data in the fields of stellar astrophysics, extragalactic astronomy, the physics of the interstellar medium, and solar system bodies. A distinctive feature of the IR camera installed on the CMO is a fairly large field of view relative to the JWST, MUSE, and a number of other telescopes, allowing for obtaining images and spectra of extended objects (with a diameter of several arcminutes) to be obtained. One of the most challenging problems in processing IR spectroscopic observations is correcting the spectrograms for the sky background. Very bright atmospheric hydroxyl lines emit at wavelengths of about 2~$\mu$m (see, for example, Fig.~6 in the paper~\cite{zheltoukhov2020}). While the sky background for point sources or sources with small angular sizes (up to 3~arcmin) can be determined directly from the object's spectrogram, extended sources, such as nebulae and galaxies with angular sizes greater than 3-4~arcmin, require additional sky spectra to be taken in the immediate vicinity of the object. Unfortunately, atmospheric hydroxyl lines change their intensity significantly in a time shorter than the exposure and reading time of a single frame (about 10~minutes; see Section~\ref{sect:lines_t}). This is a consequence of the fact that the CMO is located in a place with not the best astroclimate for IR observations: the median value of the atmospheric precipitable water vapor (PWV) is 7.7~mm and is subject to rapid changes (ground-based observatories in Chile and Hawaii, where specialized IR telescopes are installed, have PWV<2~mm).

Our survey of regions of massive star formation in the optical, infrared, and millimeter ranges (OPTIMus)~\cite{kirsanova2026} includes, among other things, infrared spectral observations. The objects of study -- regions of modern star formation in the Galaxy -- are natural physico-chemical laboratories where a wide variety of ionization states, physical conditions, and velocities are observed. These regions have angular dimensions reaching tens of arcminutes, and the H$_2$ transition emission lines of particular interest to us in these regions are many times weaker than the atmospheric hydroxyl lines; many of them lie at the peaks or wings of atmospheric lines. This required the development of a special technique for determining, accounting for, and subtracting the sky background spectrum.

As a sample, we used the long-slit spectra of the NGC~7538 (S158) region. This region, containing an ionizing star of spectral type O5-O6~V (the IRS6 source), is located at a distance of 2.8~kpc and has an angular diameter of about 10~arcmin~\cite{ojha2004}. Previously, based on observations obtained at the CMO, we had already carried out IR photometric studies of the NGC~7538 region in the $H$ band and narrow-band filters centered on the Br$\gamma$ ($\lambda=2.17$~$\mu$m), H$_2$ (2.12~$\mu$m) and [Fe\,II] (1.64~$\mu$m) lines~\cite{kirsanova2023}.

\section{Observations}

\begin{figure}[h]
\center{\includegraphics[width=0.4\linewidth]{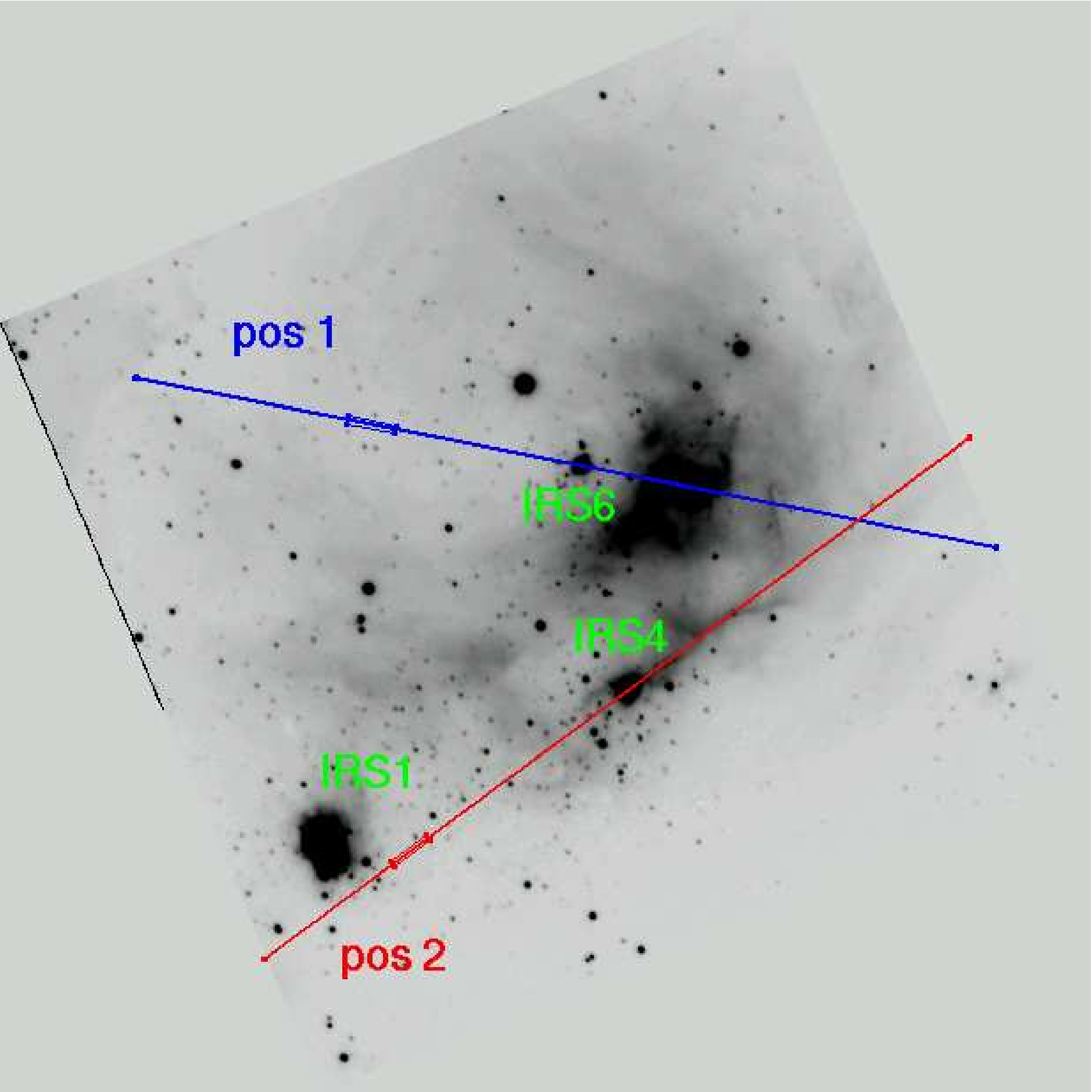}}
\caption{\label{fig:map1}Br$\gamma$ image of NGC~7538 from~\cite{kirsanova2023} with plotted positions of slits~1 and 2 (pos~1 and 2). Bold sections enclose the regions where the analysis of line and continuum brightness variations was performed in Section~\ref{sect:lines_t} and their correction in Section~\ref{sect:method}. The size of the image in Br$\gamma$ is $5.9\times5.9$~arcmin$^2$, the length of the slits is 268~arcsec. North is at the top, east is at the left.}
\end{figure}

Long-slit spectroscopic observations of NGC~7538 were carried out by the authors on February~15, 2022 (slit~1 position, centered on IRS6, position angle (PA) 78.8\hbox{$^\circ$}) and September~26, 2022 (slit~2 position, centered on IRS4, PA$=126.5$\hbox{$^\circ$}; Fig.~\ref{fig:map1}). Immediately after the observations of the object, a background area in the immediate vicinity of NGC~7538 was imaged with the same PA slit position and the same exposure. The image seeing during observations at slit~1 was 2.3~arcsec, and at slit~2 was 1.4~arcsec.

The AstroNIRCam camera receiver is a HAWAII-2RG matrix of the $2048\times2048$ size with a pixel size of 18~$\mu$m, which corresponds to 0.27~arcsec. In the spectral mode, only a part of the detector surface is used. The slit size is 268~arcsec~$\times$~0.9~arcsec. With this slit width, a spectral resolving power of $R=1200$ is achieved. To isolate the spectral range of wavelengths and the corresponding order of grism dispersion, a $K$ filter from the set of standard broadband filters was used. This made it possible to obtain spectra in the range of 2.04-2.35~$\mu$m with a dispersion of 4.6\AA/pixel.

\begin{figure}[h]
\center{\includegraphics[width=0.9\linewidth]{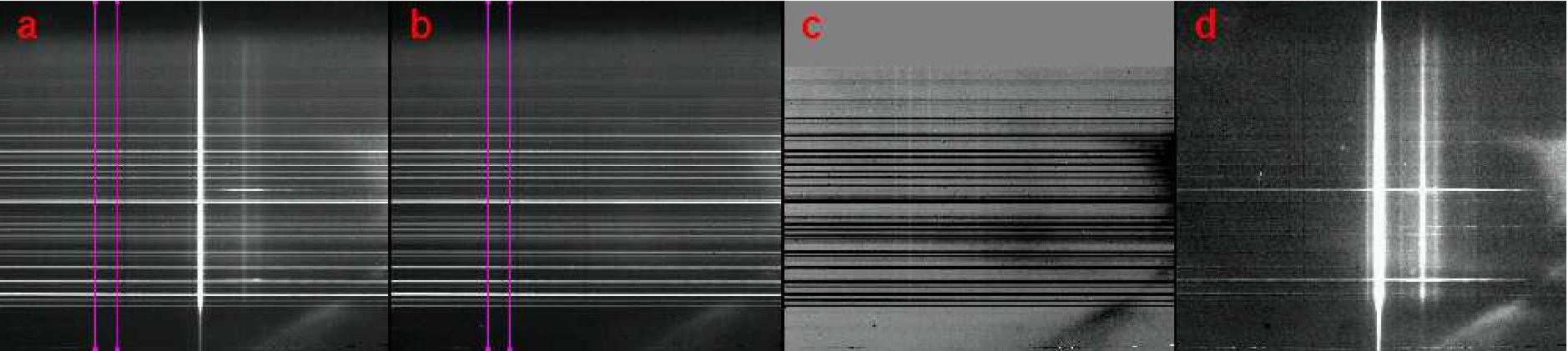}}
\caption{\label{fig:map2}The summed and normalized to the exposure time spectrograms for slit~1 of the object (a) and the sky area (b); the difference between the sky background spectrogram corrected according to the technique described in Section~\ref{sect:method} and the uncorrected background spectrogram (c); the spectrogram of the object after subtraction by the corrected sky background spectrogram (d). The vertical lines in images (a, b) delimit the area in which the analysis of line and continuum brightness changes in Section~\ref{sect:lines_t} and their correction in Section~\ref{sect:method} were performed. The coordinates along the slit are located horizontally, the wavelengths are located vertically.}
\end{figure}

\begin{figure}[h]
\center{\includegraphics[width=0.9\linewidth]{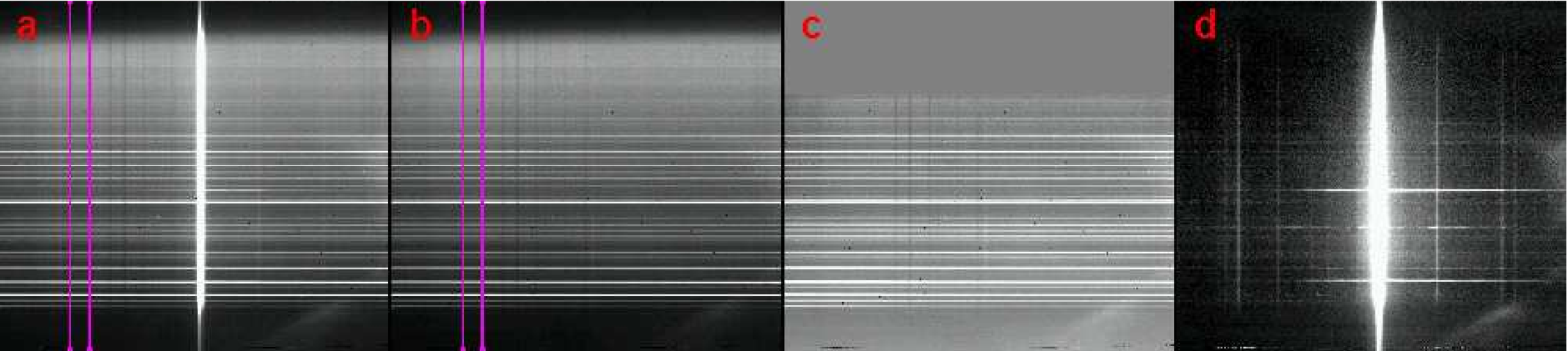}}
\caption{\label{fig:map3}The same as in Fig.~\ref{fig:map2}, but for slit~2.}
\end{figure}

During observations on February~15, 2022, four spectrograms of NGC~7538 were obtained (one with an integral exposure time of 292.9~s and three with an exposure of 596.7~s, hereinafter designated obj~1-4) and two spectrograms of the sky background (each with an exposure of 596.7~s; sky~1-2). On September~26, 2022, three spectrograms of the object (obj~1-3) and two of the sky background (sky~1-2) were obtained, each with an exposure of 592.9~s. Immediately after observing the sky background, spectra of standard stars were obtained (three spectrograms of HIP22590 in February with an exposure of 97.6~s each and three spectrograms of HIP117450 in September with an exposure of 192.9~s each), taken on nearby air masses. Primary data processing was carried out according to the standard method described in the paper~\cite{zheltoukhov2020}. The output data after processing are FITS files of $992\times900$~pixels in size (Fig.~\ref{fig:map2}, \ref{fig:map3}).

\section{Changes in brightness of sky lines and continuum level over time}
\label{sect:lines_t}

\begin{figure}[h]
\center{\includegraphics[width=0.9\linewidth]{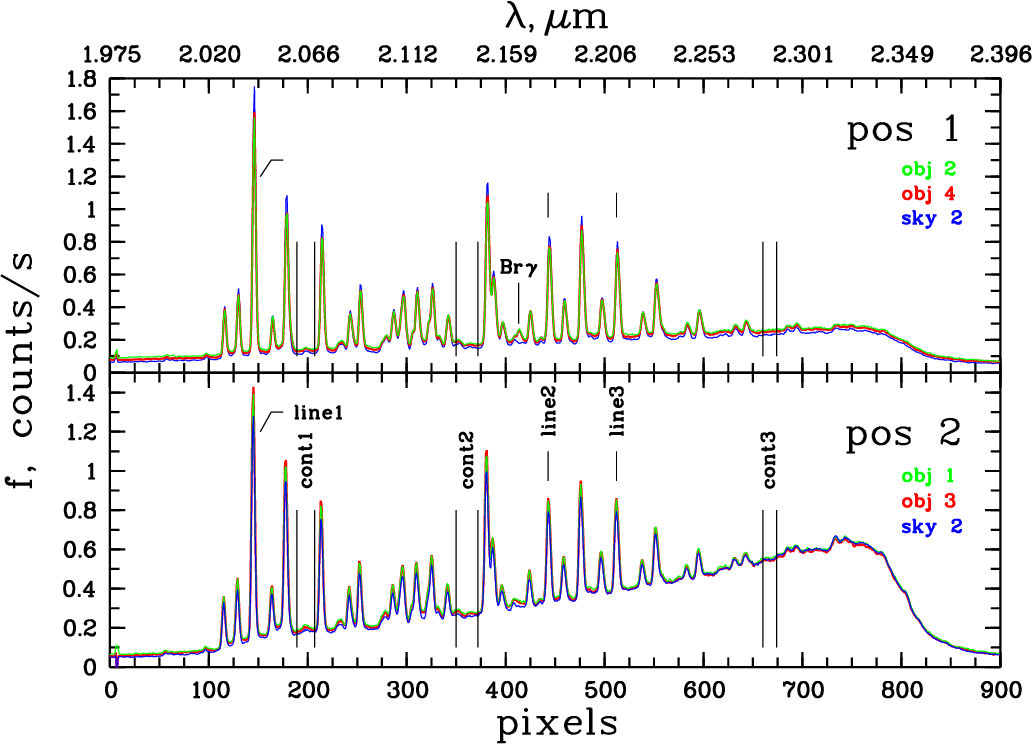}}
\caption{\label{fig:fig1}Spectra of the object (the second and fourth) and the second spectrum of the sky background at slit position~1 (green, red, and blue curves, respectively; upper graph), and spectra of the object (the first and third) and the second spectrum of the sky background at slit position~2 (green, red, and blue curves, respectively; lower graph) for the area shown in Figs.~\ref{fig:map1}-\ref{fig:map3}. The difference between the images is $\sim20$~minutes. The arrows indicate the hydroxyl lines 2.0413 (line~1), 2.1802 (line~2), and 2.2126~$\mu$m (line~3); The horizontal lines show the areas of continuum at $\lambda\sim2.06$~$\mu$m ($\sim200$~pixels; cont~1), $\lambda\sim2.14$~$\mu$m ($\sim370$~pixels; cont~2) and $\lambda\sim2.29$~$\mu$m ($\sim670$~pixels; cont~3), will be analyzed in Fig.~\ref{fig:fig2}.}
\end{figure}

To measure changes in atmospheric line brightness and continuum levels over time, we used spectrograms of the object and background regions. For this purpose, we selected the faintest regions of NGC~7538 along slits~1 and 2, where the contribution of the object's emission lines to the spectrum is minimal (see bold bars in Fig.~\ref{fig:map1} and vertical lines in Figs.~\ref{fig:map2} and \ref{fig:map3}). To obtain homogeneous data, we selected the same slit regions for the spectrograms of the object and the background (separately for slit positions~1 and 2). The first spectrum of NGC~7538 (obj~1) at slit position~1, obtained with a different exposure, was excluded from the analysis.

For each pixel across the slit (along the $\lambda$; $Y$-axis in Figs.~\ref{fig:map2}, \ref{fig:map3}), the median fluxes were calculated, normalized to the exposure time in seconds. The resulting spectra are presented in Fig.~\ref{fig:fig1}.

We selected three bright atmospheric lines (line~1-3) and three continuum spectrum areas (cont~1-3) at different $\lambda$ that do not overlap with the object's bright emission lines (Fig.~\ref{fig:fig1}). For the hydroxyl lines, we measured the integrated fluxes and line profile brightness maxima (subtracting the continuum around the lines), and for the continuum spectrum areas, we measured their average values.

By fitting the hydroxyl line profiles with Gaussians, we found that over the course of an observation set (2-3~hours), the pixel grid wavelengths $\lambda_0$ and full width at half maximum (FWHM) $\sigma$ of the sky lines generally remain stable: the relative variations $\Delta\lambda_0/\sigma$ for the three sky lines shown in Fig.~\ref{fig:fig1} do not exceed $2.5\%$ (0.5\AA\,or 0.1 pixels in absolute units) for both sets, and the $\Delta\sigma/\sigma$ does not exceed $2.5\%$ (0.5\AA\,or 0.1 pixels) for a typical $\sigma\sim4$~pixel (16-20\AA). Possible exceptions are discussed in Section~\ref{sect:method}.

Given the stability of the hydroxyl line parameters, we did not consider the integrated fluxes of the lines, as the temporal variations in the line profile along the pixel grid (i.e., for each value of $\lambda$) are only important for our task.

\begin{figure}[h]
\center{\includegraphics[width=0.4\linewidth]{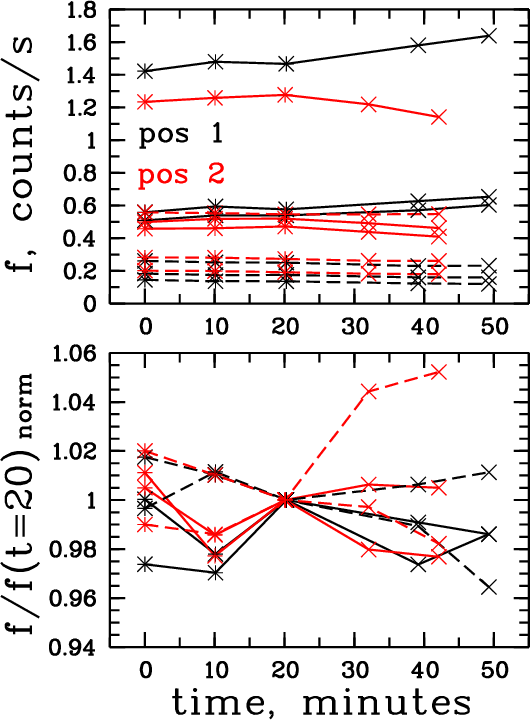}}
\caption{\label{fig:fig2}Changes in the maximum brightness levels of sky lines 1-3 $f^{\rm max}_{\rm line}$ (solid lines) and the average values of the continuum level in the spectral areas~1-3 $f_{\rm cont}$ (dashed lines) with time for observations with slit positions~1 (black) and~2 (red). Asterisks indicate values for the object's spectra, and diagonal crosses indicate values for the sky's spectra. The top graph shows the absolute flux values, the bottom graph shows the ratios $f^{\rm max}_{\rm line~1}/f^{\rm max}_{\rm line~2}$, $f^{\rm max}_{\rm line~3}/f^{\rm max}_{\rm line~2}$, $f_{\rm cont~1}/f_{\rm cont~2}$, and $f_{\rm cont~3}/f_{\rm cont~2}$, normalized to the corresponding flux ratios for the last exposure of the object, corresponding to the time $\approx20$~minutes. The time $t=0$ corresponds to the recording of the first exposure file (obj~2 for slit~1 and obj~1 for slit~2).}
\end{figure}

Fig.~\ref{fig:fig1} shows the obtained spectra for three exposures at each slit position. As can be seen from the figure, the sky lines vary significantly in brightness from spectrum to spectrum. The change in the maximum intensity of the brightest hydroxyl line (line~1) over the observation period ($\approx50$~minutes, see Fig.~\ref{fig:fig2}) reaches $0.15-0.2$~counts per second, corresponding to relative changes of $10-15\%$. For comparison, the brightest line in the spectrum of the nebula NGC~7538, Br$\gamma$ ($\lambda=2.1661$~$\mu$m), is not blended with atmospheric lines and is visible in the spectra as a weak peak at $X\approx415$~pixels (see the difference between the spectra of the object (obj) and the sky (sky) in Fig.~\ref{fig:fig1}).

The ratio of intensities (brightness maxima) of atmospheric lines remains quite stable, changing within $\pm1\%$ (see the lower graph in Fig.~\ref{fig:fig2}).

The change in the continuum level with time in absolute units is extremely small, within $2\%$ (Fig.~\ref{fig:fig2}). In relative units, variations in the continuum level and its ratio in different parts of the spectrum roughly correspond to changes in the maximum sky brightness and their ratios. The only exception is the increase in the continuum level in the long-wavelength region of the spectrum during observations at slit position~2 (see the lower graph in Fig.~\ref{fig:fig2}), and the continuum level in the spectrograms of the background area is even slightly higher than the level in the spectrograms of the object (see the $X\approx700$~pixels ($\lambda\approx2.30$~$\mu$m) region in the lower graph in Fig.~\ref{fig:fig1}). A similar case: an increase in the slope of the continuum -- is typical for observations during which the telescope temperature changes. In particular, during the observation period in the position of slit~2 of the object and the sky background, the temperature of the telescope tube and mirror increased by $0.2{\hbox{$^\circ$}}$C.

Note that, unlike variations in the brightness of atmospheric lines, variations in the continuum do not affect our primary goal is the measuring the intensity of the object's emission lines. In the OPTIMus we study the properties of the gas-dust environment in regions where massive stars form. The continuum level in the spectra of ''pure'' (without admixture of stars) interstellar medium is extremely small; the equivalent widths of emission lines in such spectra are uninformative. The continuum level in the spectra of stars in star-forming regions, especially in the spectra of bright sources (IRS4, IRS6, etc.), is quite high relative to variations in the continuum during the observation set and allows us to obtain stellar spectra with a high signal-to-noise ratio (see the spectra of stars in the left and right images of Fig.~\ref{fig:map2}, \ref{fig:map3}).

\section{Method of correcting spectra for atmospheric lines}
\label{sect:method}

In Section~\ref{sect:lines_t}, it was shown that the intensities of sky lines vary significantly from frame to frame during the observation time, while their ratio, as well as the continuum level, varies very little. Thus, to correct the atmospheric hydroxyl lines, it is necessary to find the ratio of the fluxes in the lines in the spectrograms of the sky background and the object. For this, we used the same regions along the slit in the spectrograms of the nebula and the background area as in the previous Section (see Figs.~\ref{fig:map1}-\ref{fig:map3}). A fundamental point is the choice of regions in which there are no stars that make a significant contribution to the continuum level. To further verify the method, we also examined spectrograms of standard stars (star) in the same regions along the slit, free from the contribution of radiation from astronomical objects to the spectrum. The contribution to the spectrum in these regions is made only by sky background radiation.

\begin{figure}[h]
\center{\includegraphics[width=0.9\linewidth]{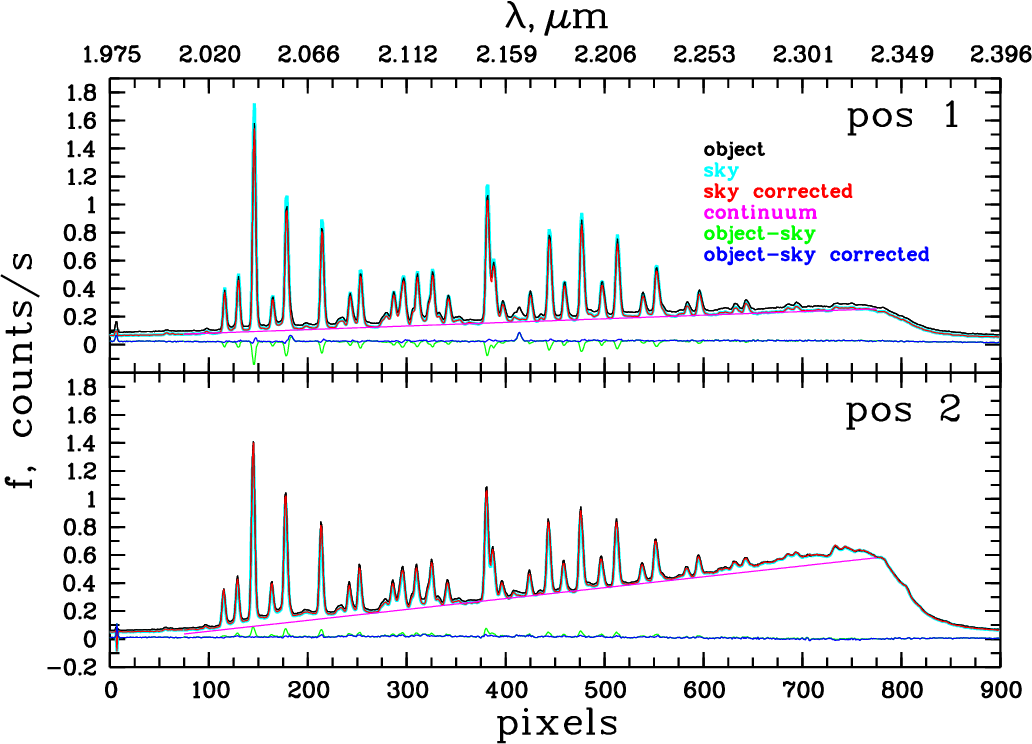}}
\caption{\label{fig:fig3}Spectra of the object (black curve), sky background (cyan), corrected sky background (red), continuum level (magenta line), difference between the spectrum of the object and the sky (green curve), and difference between the spectrum of the object and the corrected sky background (blue) for slit positions~1 (top) and~2 (bottom).}
\end{figure}

Unlike Section~\ref{sect:lines_t}, here we summarized the images of the object spectrum, the exposure spectra of the background area, and the spectra of area from the spectrograms of standard stars for each slit position. Next, for each pixel across the slit, the median fluxes were determined and normalized to the exposure time in seconds. The resulting summarized spectra of the objects and background regions (black and cyan curves, respectively) are shown in Fig.~\ref{fig:fig3}.

We first examined the correspondence between the wavelengths $\lambda_0$ on the pixel grid in the spectra of the object, the sky background, and the area from the standard star spectrogram for each slit position. As shown in Section~\ref{sect:lines_t}, the deviations in $\lambda_0$ did not exceed 0.1~pixels, with the exception of the standard star spectrum for slit position~1. This spectrogram was shifted along the pixel grid by 0.2~pixels to align with the sky background spectrogram for slit~1. The amplitudes of the sky lines changed in relative units by $\approx2\%$ (0.02~counts/s for the brightest lines) when shifted.

In the first step, we found the continuum level for the sky background spectrum. Considering that hydroxyl lines are absent at the edges of the spectrum and the transmission of the $K$ filter used in the observations, which cuts the spectral range of wavelengths, drops sharply (see Figs.~1, 5 in~\cite{zheltoukhov2020}, we found the continuum level in the wavelength range from 2.01 to 2.34~$\mu$m ($X=75-775$~pixels; Fig.~\ref{fig:fig3}). In this region, the continuum level can be considered to increase linearly with increasing lambda (i.e., $X$). At the edges of the spectrum, outside the $\lambda=2.01-2.34$~$\mu$m region, the $K$ filter sensitivity drops sharply, and atmospheric hydroxyl lines are absent. The obtained dependencies for the continuum level are shown by magenta lines in Fig.~\ref{fig:fig3}.

\begin{figure}[h]
\center{\includegraphics[width=0.9\linewidth]{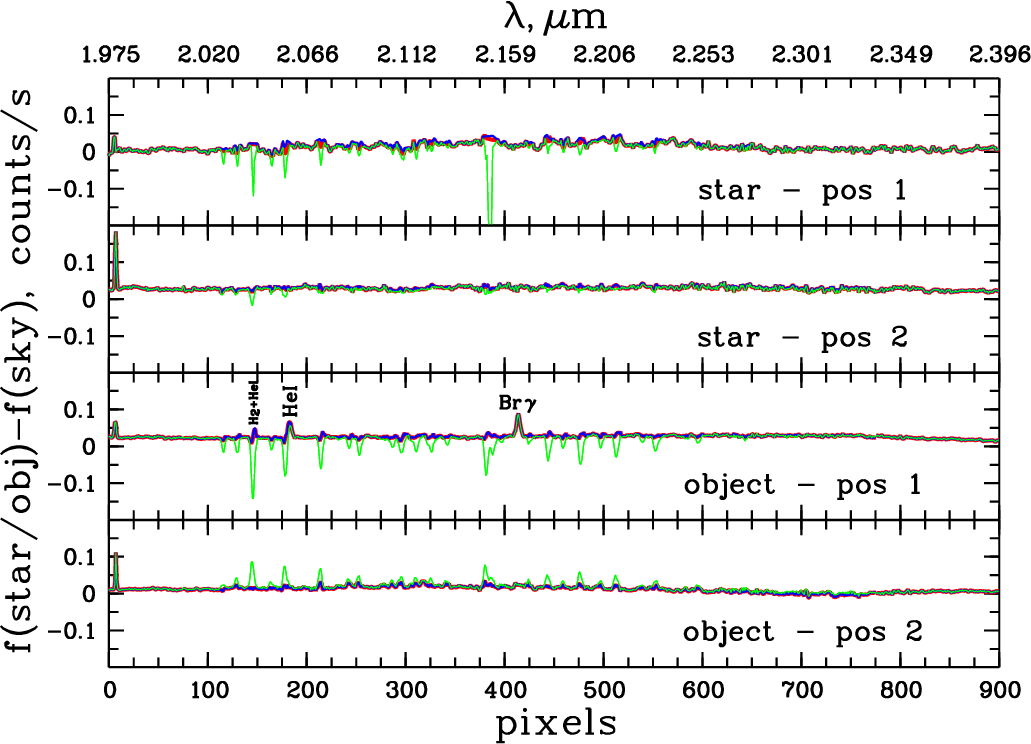}}
\caption{\label{fig:fig3b}The difference between the spectrum of the object (standard star area) and the sky (green curves), the difference between the spectrum of the object (standard star area) and the sky background, corrected according to the approximation by the lower envelope of the point cloud (blue curves) and corrected according to the approximation by the entire point cloud (red curves) for slit positions~1 and 2.}
\end{figure}

In the second step, the continuum $f({\rm cont})$ was subtracted from the sky background spectrum $f({\rm sky})$. At the same time, we subtracted the sky spectrum (ignoring the continuum level) from the spectra of the object and the standard star area. The resulting spectral difference $f({\rm obj})-f({\rm sky})$, $f({\rm star})-f({\rm sky})$ for each pixel $X$ is shown in green in Figs.~\ref{fig:fig3}, \ref{fig:fig3b}.

\begin{figure}[h]
\center{\includegraphics[width=0.9\linewidth]{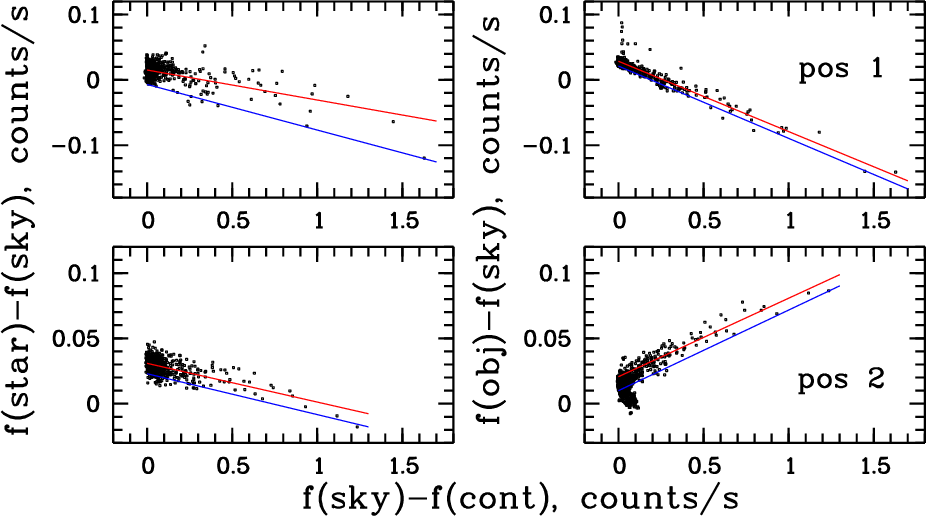}}
\caption{\label{fig:fig4}The dependence between the flux difference for each pixel ($X$-axis in Fig.~\ref{fig:fig3}, \ref{fig:fig3b}) in the spectrum of the standard star area $f({\rm star})$ (left), object $f({\rm obj})$ (right) and background area $f({\rm sky})$, and the flux in the spectrum of the sky background area after continuum subtraction $f({\rm cont})$ for observations with slit positions~1 (top) and 2 (bottom). The red and blue lines are the obtained estimates of the linear regression~(\ref{eq:eq1}) of the point cloud, and by the lower envelope of the point cloud, respectively.}
\end{figure}

Further analysis was carried out using the diagrams $(f({\rm star})-f({\rm sky})) - (f({\rm sky})-f({\rm cont}))$, $(f({\rm obj})-f({\rm sky})) - (f({\rm sky})-f({\rm cont}))$ (Fig.~\ref{fig:fig4}).

The graphs in Fig.~\ref{fig:fig4} show that systematic changes are observed in the bright atmospheric lines (for the values of $f({\rm sky})-f({\rm cont})>0.1$~counts/s). It decreases with increasing $f({\rm sky})-f({\rm cont})$ in the case when the hydroxyl lines are brighter in the spectrogram of the background area, and increases in the opposite case. The linearity of the dependence at $f({\rm sky})-f({\rm cont})>0.1$~counts/s reflects the fact that their ratio does not change in the spectrum of the sky background relative the spectrum of the object (the standard star area). At zero values of $f({\rm sky})-f({\rm cont})$ the difference $f({\rm obj})-f({\rm sky})=0.01-0.02$~counts/s; this is a consequence of the fact that the level of the continuum of the object spectrum systematically exceeds the level of the continuum of the sky background spectrum (see the spectra in Figs.~\ref{fig:fig1} and \ref{fig:fig3} for both slits). For the spectrum obtained at slit position~2, the difference $f({\rm obj})-f({\rm sky})$ at small $f({\rm sky})-f({\rm cont})$ varies in the range of $-0.005\div0.025$~counts/s. The reason for this is the different slope of the continuum for the object and the sky background spectra; in the long-wave part, the continuum level of the spectrum of the sky area exceeds the continuum level of the spectrum of the object. This is especially clearly seen in the lower graph of Fig.~\ref{fig:fig1} at $\lambda>2.30$~$\mu$m ($X>700$~pixels).

The scatter of points in the $f({\rm star})-f({\rm sky})$ diagram for the spectra of standard star fields turned out to be noticeably larger than in the $f({\rm obj})-f({\rm sky})$ diagram for the object's spectra (Fig.~\ref{fig:fig4}). The reason for this is the relatively short exposure time of the standard star spectra and, as a consequence, a smaller signal-to-noise ratio for them. We also note the non-zero difference $f({\rm star})-f({\rm sky})$ at zero values of $f({\rm sky})-f({\rm cont})$, reflecting variations in the level of the continuum over time.

At the next stage, the linear regression is calculated using the points on the diagrams $(f({\rm star})-f({\rm sky})) - (f({\rm sky})-f({\rm cont}))$ and $(f({\rm obj})-f({\rm sky})) - (f({\rm sky})-f({\rm cont}))$:
\begin{equation}\label{eq:eq1}
f({\rm star/obj})-f({\rm sky})=\alpha(f({\rm sky})-f({\rm cont}))+\beta.
\end{equation}
We considered two linear approximation options: one for the entire point cloud in the diagrams (excluding points with $f({\rm sky})-f({\rm cont})<0.15$~counts/s for the object's spectra; red lines in Fig.~\ref{fig:fig4}) and one for the lower envelope of the point cloud (blue lines).

The option with a linear approximation by the lower envelope of the point cloud was carried out because for the object's spectrum, higher values of $f({\rm obj})-f({\rm sky})$ for a given $f({\rm sky})-f({\rm cont})$ can be caused by the contribution of nebula emission lines.

The relative error of the linear approximation $\Delta\alpha/|\alpha|$ was $4-6\%$; the difference between the regression coefficients obtained for the lower envelope of the point cloud and for the entire point cloud lies within this error for both slit positions. The only exception was the dependence for the standard star area at slit position~1 (see the upper left graph in Fig.~\ref{fig:fig4}).

Next, each pixel of the two-dimensional spectrogram of the sky background is multiplied by the coefficient $\alpha+1$, and the continuous spectrum is added to it:
\begin{equation}\label{eq:eq2}
f({\rm sky\,corr})(X,Y)=(\alpha+1)f({\rm sky})(X,Y)+f({\rm cont})(Y).
\end{equation}
The spectrogram of the sky background $f({\rm sky\,corr})$, corrected according to equation~(\ref{eq:eq2}), is subtracted from the spectrogram of the object $f({\rm obj})$ or the spectrogram of area of the standard star $f({\rm star})$.

The difference between the spectrum of the object (the standard star area) and the corrected sky background for both slit positions is shown in Fig.~\ref{fig:fig3b}, where the blue curve corresponds to the corrected sky background according to the approximation based on the lower envelope of the point cloud, and the red curve corresponds to the approximation based on the entire point cloud. As can be seen from the figure, both approximations give virtually identical results and they correct the spectrum for the sky lines (green curve) well. The profile $f({\rm star})-f({\rm sky})$ in the upper graphs of Fig.~\ref{fig:fig3b} corresponds to the difference between the levels of the continuum of the standard star area spectrum and the sky background.

In general case, to correct sky lines when observing extended objects, we recommend using an approximation based on a point cloud lower envelope, performing a linear regression over the entire point cloud for verification.

The resulting nebula spectrograms, after subtracting the sky background spectrogram corrected according to the approximation using a point cloud lower envelope, are shown in Figs.~\ref{fig:map2}d and \ref{fig:map3}d for slit positions~1 and 2, respectively. In Fig.~\ref{fig:fig3}, the corrected sky background spectra are shown by red, while the object spectra after subtracting the corrected sky background spectrum are shown by blue.

Outside the emission lines, the continuum level of the spectrum $f({\rm obj})-f({\rm sky\,corr})$ coincides with the continuum level $f({\rm obj})-f({\rm sky})$ with an accuracy of better than $10\%$ (cf. the blue and green curves in Fig.~\ref{fig:fig3} and the profiles in Fig.~\ref{fig:fig3b}). Thus, we are able to correct the sky background spectrum with the correct accounting of the continuum level.

Sky background spectrum correction is performed for the central region of the spectrum. For the short-wavelength ($\lambda<2.01$~$\mu$m; $X<75$) and long-wavelength ($\lambda>2.34$~$\mu$m; $X>775$) regions, the sky background spectrum is subtracted from the object's spectrum without any corrections.

The difference between the corrected ($f({\rm sky\,corr})$) and original ($f({\rm sky})$) sky background spectrograms is shown in Figs.~\ref{fig:map2}c and \ref{fig:map3}c for slit positions~1 and 2, respectively. The spectra of regions of the NGC~7538 nebula after sky background subtraction are shown in Fig.~\ref{fig:fig3}.

Comparing the spectra of the object after subtracting the corrected and original spectrum of the sky background (blue and green lines in Fig.~\ref{fig:fig3}), it can be noted that: i) we were able to exclude the influence of bright atmospheric hydroxyl lines, ii) against the background of atmospheric lines, we were able to distinguish weak emission lines even in a region of the nebula with a low density gas-dust environment. This is especially noticeable in the corrected spectrum of the nebula at slit position~1 (Figs.~\ref{fig:fig3}, \ref{fig:fig3b}), where the following lines stand out: H$_2\,\lambda2.0418$~$\mu$m+He~I\,$\lambda2.0431$~$\mu$m ($X=145$~pixels), He~I\,$\lambda2.0587$~$\mu$m ($X=180$~pixels), Br$\gamma$\,$\lambda2.1661$~$\mu$m ($X=415$~pixels).

\section{Discussion and conclusions}

Our analysis of spectrograms revealed that atmospheric line intensities vary significantly ($\sim10\%$) over the course of an observation set; however, the sky line ratio, as well as the spectral continuum level, vary very little (within $\sim1\%$).

We have developed a technique for correctly recovering the intensities of bright (relative to the object's lines) atmospheric hydroxyl lines obtained during spectral observations of extended objects whose angular dimensions exceed the spectrograph slit length.

The proposed sky background correction technique enables accurate measurements of H$_2$ and some other (He~I, H~I, [Fe~III]) line fluxes and the derivation of interstellar medium parameters: the temperature and column density of molecular hydrogen in star-forming regions. Preliminary results have shown that for the brightest regions of the NGC~7538 nebula it is possible to study more subtle effects, such as the estimation of the para- to ortho-hydrogen ratio~\cite{puxley2000}.

\section*{ACKNOWLEDGMENTS}
The authors thank the reviewers for their helpful comments. The work was carried out using equipment acquired through the Lomonosov Moscow State University Development Program.

\section*{FUNDING}
The study was conducted under the state assignment of Lomonosov Moscow State University (observations, data processing, development and description of the methodology, and writing the text) and the state assignment of the Institute of Astronomy of the Russian Academy of Sciences (project statement and text editing).

\section*{CONFLICT OF INTEREST}

The authors of this work declare that they have no conflicts of interest.



\end{document}